\documentclass[]{iopart}
\usepackage{iopams}
\usepackage{setstack}
\usepackage[]{graphicx}
\usepackage{color}
\usepackage[latin1]{inputenc}
\usepackage[english]{babel}
\usepackage{dsfont}

\newcommand{\colr}{}

\newcommand{\cU}{{\cal{E}}}
\newcommand{\cF}{{\cal{F}}}
\newcommand{\rT}{{\rm{T}}}

\begin{document}

\title[Stray capacitances in the watt balance operation]{Stray capacitances in the watt balance operation: electrostatic forces}
\author{D Quagliotti and G Mana}
\address{INRIM -- Istituto Nazionale di Ricerca Metrologica, str.\ delle Cacce 91, 10135 Torino, Italy}

\begin{abstract}
In a watt balance, stray capacitances exist between the coil and the magnet. Since the electric current flowing in the coil originates a difference between the coil and magnet electric-potentials, their electrostatic interactions must be taken into account. This paper reports the results of a finite element analysis of the forces acting on the coil.
\end{abstract}

\submitto{Metrologia}
%06.20.Jr Determination of fundamental constants
%03.50.De Classical electromagnetism, Maxwell equations

\ead{d.quagliotti@inrim.it}

\section{Introduction}
A watt balance virtually compares the mechanical and the electric powers produced by the motion of a mass in the Earth gravitational field and by the motion of the supporting coil in a magnetic field \cite{Kibble:1976,Kibble:1987}. The vertical force acting on a coil linking the magnetic flux $\Phi$ and carrying the electric current $I$ is $F=I\partial_z \Phi$, where $\partial_z$ indicates the derivative along the vertical. The electromotive force along the same coil moving vertically at velocity $u_z$ is $\cU=-u_z\partial_z \Phi$. If $F$ counterbalances the weight $-mg$ of a mass in the gravitational field, by combining these equations and eliminating the geometric factor $\partial_z \Phi$ we obtain $mgu+\cU I=0$. This equation relates mechanical and electric powers and allows either $m$ to be determined in terms of electric quantities or the Planck constant to be determined in terms of mechanical quantities \cite{Mana:2012,Steiner:2013}.

An assumption in this analysis is that no additional force acts on the coil. However, stray capacitances exist between the coil and the magnet and, since the electric current flowing in the coil originates a difference between the coil and magnet electric-potentials, electrostatic forces act on the coil. To support and to complement the watt-balance measurements of the Planck constant, we extended previous investigations on the coil-field interaction \cite{Sasso:2013,Mana:2013} by quantifying these forces.

{\colr In the experiments up to now completed, these forces have been implicitly assumed uninfluential, but, to our knowledge, no detailed study has been carried out to support this statement. Owing to the extreme accuracy of the Planck constant measurements, to gain confidence in the uncertainty of the no-effect statement, an equivalently accurate analysis is necessary. We have been also motivated by the spread -- the order of magnitude of which is 100~nW/W, to be compared with a targeted uncertainty of 10~nW/W -- of the Planck constant values reported by the International Avogadro Coordination (IAC) \cite{Andreas:2011a,Andreas:2011b}, the National Institute of Standards and Technology (NIST-USA) \cite{Steiner:2005,Steiner:2007}, the Swiss Federal Office of Metrology (METAS-Switzerland) \cite{Eichenberger:2011}, the National Physical Laboratory (NPL-UK) \cite{Robinson:2012a}, and the National Research Council (NRC-Canada) \cite{Steele:2012}. Our study demonstrates that an accurate calculation of these minute forces is possible and that their effect is actually irrelevant, thus confirming the no-effect assumption.}
\begin{figure}
\centering
\includegraphics[width=65mm]{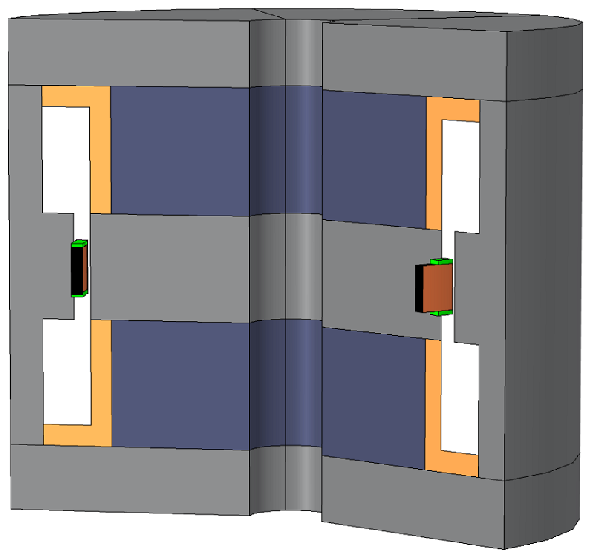}
\includegraphics[width=40mm]{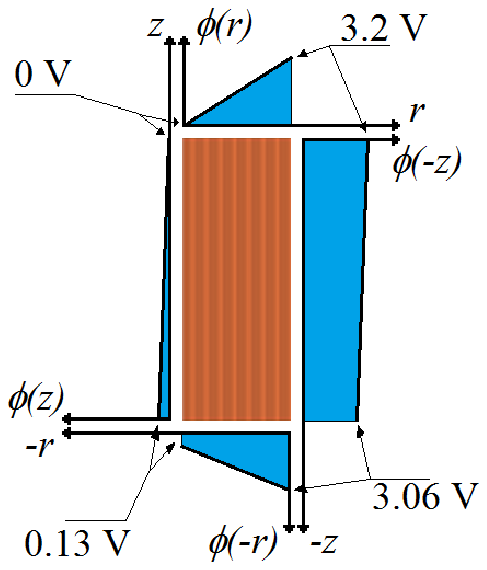}
\caption{Model of the coil-magnet system. The coil is a metallic toroid having a rectangular cross section. It is supported by two dielectric rings (green) and enclosed into a toroidal cavity (white). The electrical potential of the cavity surfaces is null. Owing to the flow of the electric current, the electric potential of the coil surfaces varies linearly as shown by the blue profiles (right).}\label{toy}
\end{figure}

\section{Coil-magnet system}
{\colr The analysis of a real watt balance system is hampered by its complexity. Therefore, since we are not running any real experiment and, consequently, we have no pretension to completeness and to carry out an analysis of the error budget of a specific experiment, we cut the system design by limiting the investigation to the coil-magnet system and by turning to the model shown in Fig.~\ref{toy}. In particular, since incremental refinements might be endless, we focused on investigating if and how the electrostatic forces can be calculated accurately enough to exclude with certainty an effect in the watt balance operation.}

We made reference to the coil-magnet system of the METAS watt balance; dimensions were taken from \cite{Baumann:2013}. The sources of the magnetic field are two permanent magnets (blue in Fig.~\ref{toy}) placed on both sides of the kernel (grey in Fig.~\ref{toy}) with the same poles facing each other. The adjustment of the magnets to the kernel and the centring of the magnets-kernel assembly to the yoke is ensured by two bronze centring rings (yellow in Fig.~\ref{toy}).

\begin{figure}
\centering
\includegraphics[width=40mm]{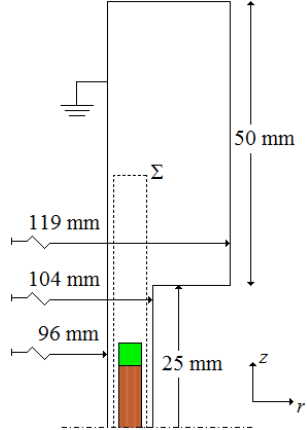}
\caption{Radial section of the upper half-domain of the Laplace equation (\ref{field:eq}). The dashed box $\Sigma$ is the section of the torus used to calculate the forces acting on the coil (brown-coloured). The coil dimensions are: inner radius 97~mm, outer radius 103~mm, height 21~mm. The height of the clamping rings (green-coloured) is 5~mm. The boundary conditions on the coil surface are shown in Fig.\ \ref{toy} (right).}\label{toy-2}
\end{figure}

A coil (brown in Fig.~\ref{toy} and \ref{toy-2}) of 1836 windings -- having mean diameter 200~mm, height 21~mm, wire diameter 250~$\mu$m, electrical resistance 458~$\Omega$, and inductance 1.23~H -- is pinched between two ceramic rings (green in Fig.~\ref{toy}) and supported by six ceramic legs mounted on a hexagonal plate below and outside the magnetic circuit (not shown in Fig.~\ref{toy}).

\section{Electrostatic forces}\label{eltforces}
\subsection{Lumped parameter model}
\begin{figure}
\centering
\includegraphics[width=75mm]{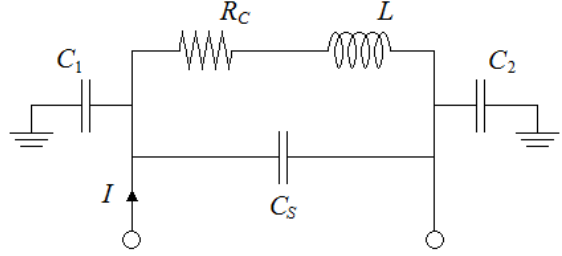}
\caption{Lumped parameter model of the coil-magnet system. $C_1$ and $C_2$ are the stray capacitances between the inner and outer coil-layers and the magnet. $R_C$ is the coil resistance.}\label{coil}
\end{figure}

Before carrying out a numerical analysis, we make an order-of-magnitude estimate of electrostatic force acting on the coil by using the lumped parameter model shown in Fig.~\ref{coil}. The stray capacitances between the coil and the magnet are modelled by a couple of capacitors, whose armatures are the inner and the outer coil-layers and the magnetic circuit. An approximate estimate of these capacitances is given by
\begin{equation}\label{cylindrical}
 C = \frac{2\pi\epsilon_0 h}{\ln(b/a)} ,
\end{equation}
where $a$ and $b$ are the inner and outer radii of the armatures and $h$ is height of their overlapping parts.

The current flowing in the coil originates a difference between the electric potentials of the coil and the magnet; if a coil end is earthed, the potential of the opposite end is $V=R_C I$, where $R_C$ is the resistance and $I$ is the current. The electric potential of the coil surfaces varies as shown in Fig.~\ref{toy} but, to carry out an order-of-magnitude calculation, we set the potentials of the inner and outer layers to the constant values of zero and $R_C I$, respectively.

The vertical coordinate $z$ of the coil centre can be varied without loosing the validity of (\ref{cylindrical}); therefore, the vertical force component is
\begin{equation}\label{e-force}
 F_z = \frac{1}{2} \left(R_C I\right)^2 \, \partial_z C .
\end{equation}
Since $\partial_z C =0$ if $h$ does not vary with $z$, the inner capacitance does not contribute to $F_z$. As regards as the outer capacitance, we must consider the parallel of the capacitances of the coil parts facing the pole shoe and yoke. By using $R_C = 458$~$\Omega$, $I=7$~mA, $a=103$~mm, and $b=104$~mm (pole shoe) or $b=119$~mm (yoke), we obtain $F_z \approx \pm~30$~nN, if the coil is partially inside the magnetic gap, and $F_z \approx 0$~nN, if it is entirely outside or inside the magnetic gap. When the coil is only slightly misplaced with respect to the central position, the force predicted by (\ref{e-force}) is zero; $F_z$ is not null only if the coil misplacement is so bad that it does not correspond to any real set-up. In this case, since a current of 7~mA generates a magnetic force of 5~N, the parasitic force would be about 6~nN/N, in relative terms.

\subsection{Homogeneous media}
The electrostatic force acting on the volume $\Omega$ of an homogeneous medium is
\begin{equation}
 \bi{F}= \int_\Omega \varrho \bi{E}\, \rmd^3 x = \int_\Omega (\bnabla^\rT \bi{D}) \bi{E}\, \rmd^3 x ,
\end{equation}
where $\varrho = \bnabla^\rT \bi{D}$ is the distribution of the free charges, $\bi{E}$ and $\bi{D}$ are the electric field and the electric displacement, respectively, $\epsilon=$ const.\ is the dielectric constant, and $\bnabla^\rT=[\partial_x, \partial_y, \partial_z]$. Here and in the following we represent vectors by column matrices, the superscript $^\rT$ indicates the transpose, and the matrix product $\bi{a}^\rT\bi{b}$ substitutes for the scalar product.

Since numerical derivatives amplify the errors, in order to avoid the calculation of $\bnabla^\rT \bi{D}$, it is convenient to represent the interaction between the charge distribution and the electric field by using the Maxwell stress tensor \cite{Jackson:1998,Schwinger:1998}. In the case of a homogeneous and isotropic medium, $\bi{D}=\epsilon\bi{E}$, we can write $(\bnabla^\rT \bi{D}) \bi{E} = \epsilon (\bnabla^\rT \bi{E}) \bi{E}$, and
\numparts\begin{equation}\label{e1}
 (\bnabla^\rT \bi{E}) \bi{E} = (\bnabla^\rT \bi{E}) \bi{E} - \bi{E}\times( \bnabla \times \bi{E} ) ,
\end{equation}
because $ \bnabla \times \bi{E} = 0$. Next, by using the identity \cite{Jackson:1998}
\begin{equation}
 \bnabla (\bi{E}^\rT\bi{E}) /2 = (\bnabla \bi{E}^\rT)\bi{E} + \bi{E}\times( \bnabla \times \bi{E} ) ,
\end{equation}
we can rewrite $(\bnabla^\rT \bi{E}) \bi{E}$ as
\begin{equation}\label{prima}
 (\bnabla^\rT \bi{E}) \bi{E} = (\bnabla^\rT \bi{E}) \bi{E}  + (\bnabla \bi{E}^\rT )\bi{E}  - \bnabla  (\bi{E}^\rT \bi{E}) /2
\end{equation}
By transposing (\ref{prima}),
\begin{equation}
 (\bnabla^\rT \bi{E}) \bi{E}^\rT = \bnabla^\rT(\bi{E}\bi{E}^\rT )- \bnabla^\rT ( \bi{E}^\rT \bi{E} )  \mathds{1} /2 .
\end{equation}\endnumparts
Therefore,
\numparts\begin{equation}\label{TE0}
 \varrho \bi{E}^\rT = \bnabla^\rT \bi{T}_E ,
\end{equation}
where
\begin{equation}\label{TE}
 \bi{T}_E = \epsilon \bi{E}\bi{E}^\rT - \frac{\epsilon E^2}{2} \mathds{1}
\end{equation}\endnumparts
is the Maxwell's stress tensor and $\mathds{1}$ is the $3\times 3$ identity matrix.

Eventually, from the divergence theorem, the force and torque acting on the volume bounded by the surface $\Sigma$ are
\numparts\begin{equation}\label{force}
 \bi{F}= \oint_\Sigma \bi{T}_E \bi{n}\, \rmd \sigma
\end{equation}
and
\begin{equation}\label{torque}
 \bi{K}= \oint_\Sigma \bi{r} \times (\bi{T}_E \bi{n})\, \rmd \sigma ,
\end{equation}\endnumparts
where $\bi{r}$ is the position vector with respect to the pivot point and $\bi{n}$ is the external unit-vector normal to integration surface.

Unfortunately, for the coil-magnet system, the $\epsilon$ discontinuity at the interfaces jeopardise the derivation of (\ref{TE0}) and the usage of (\ref{force}-$b$) to find the force acting on the coil cannot be justified. Therefore, in the next section, we briefly recall the theoretical framework necessary to validate the usage of (\ref{force}-$b$).

\subsection{Inhomogeneous media}\label{s:stress}
To find the electrostatic forces in an inhomogeneous medium is a quite complex problem. We refer to the solution given in \cite{Landau:1960}, which is here outlined for the reader convenience. Detailed and comprehensive treatments of the volume and surface forces and of the mechanical stresses in electrically polarised media can be also found in \cite{Stratton:1941,Bobbio:1999}.

It is well known that the local forces per unit volume acting on a continuous medium can be written in terms of the Cauchy stress tensor $\bi{T}$ as $\bnabla^\rT \bi{T}$. Hence, the total force acting on a part of the medium having volume $\Omega$,
\numparts \begin{equation}\label{t:force}
 \bi{F} = \int_\Omega \bnabla^\rT \bi{T}\, \rmd^3 x = \oint_\Sigma \bi{T}\bi{n}\, \rmd\sigma ,
\end{equation}
can be reduced to the integration of the forces acting on its bounding surface $\Sigma$, where $\bi{n}$ is the unit outwards-vector normal to the surface element $\rmd\sigma$. Similarly, the total torque is
\begin{equation}\label{t:torque}
 \bi{K} = \oint_\Sigma \bi{r} \times (\bi{T}\bi{n})\, \rmd\sigma .
\end{equation}\endnumparts

In order to calculate the Cauchy stress tensor, let us consider a deformation of the infinitesimal volume element $\rmd^3 x$ -- where the $z$ axis is chosen locally parallel to $\bi{E}$ -- consisting of an homogeneous isothermal virtual displacement of the top $x$-$y$ face by $\bxi \rmd z$, with the electric potential unchanged. The medium inside the deformed volume exerts the $-\bi{T} \bi{n}\, \rmd\sigma$ force on the displaced face and the work done by this force,
\begin{equation}\label{eq0}
 (\bxi^\rT \bi{T} \bi{n})\, \rmd z \rmd\sigma = ( \bxi^\rT \cF \bi{n} + \Delta \cF )\, \rmd z \rmd\sigma ,
\end{equation}
is opposite to the variation of $\cF\rmd z \rmd\sigma$, where
\begin{equation}\label{free}
 \cF = \cF_0(\Theta,\rho) - \epsilon E^2/2
\end{equation}
is the the Helmholtz free energy per unit volume, the thermodynamics variables are the temperature $\Theta$, density $\rho$, and electric field $\bi{E}$, $\cF_0$ is the Helmholtz free energy per unit volume when $\bi{E}=0$, $\epsilon$ is the dielectric constant of the medium, and the constitutive relationship $\bi{D}=\epsilon\bi{E}$ and isotropy have been assumed. By carrying out the calculations of $\Delta\cF$, we obtain \cite{Landau:1960}
\begin{equation}\label{eq1}
 \bxi^\rT \bi{T} \bi{n} = \bxi^\rT \big[ \cF - \rho (\partial_\rho \cF) + \epsilon \bi{E}\bi{E}^\rT \big] \bi{n} .
\end{equation}
For the sake of simplicity, when $\bi{E}=0$, we consider only the scalar part of the stress tensor, which is associated to the pressure $P_0=\rho(\partial_\rho \cF_0)-\cF_0$ and corresponds to model the medium as a fluid. Hence, by using (\ref{free}) in (\ref{eq1}), we obtain
\begin{equation}\label{stress}
 \bi{T} = \bi{T}_0 + \epsilon \bi{E}\bi{E}^\rT - \frac{\epsilon - \rho(\partial_\rho \epsilon)}{2} E^2 \mathbf{\mathds{1}}
 = \bi{T}_E + \bi{T}_0 + \frac{1}{2} \rho E^2 (\partial_\rho \epsilon) \mathbf{\mathds{1}} ,
\end{equation}
where $\bi{T}_E$ is given by (\ref{TE}) and $\bi{T}_0$ is the stress tensor in the absence of the electric field. With the fluid-medium approximation, $\bi{T}_0 = -P_0 \mathbf{\mathds{1}}$; but, if the $\epsilon$ anisotropy induced by elastic deformations is neglected, (\ref{stress}) holds also in the general case.

In the absence of free charges and of temperature and density gradients, since $\bnabla\epsilon=0$ and, consequently, $\bnabla \bi{T}_E=0$, the condition of mechanical equilibrium is
\begin{equation}\label{equilibrium}
 \bnabla^\rT \bi{T} = \bnabla^\rT \big[ \bi{T}_0 + \frac{1}{2} \rho E^2 (\partial_\rho \epsilon) \mathbf{\mathds{1}} \big] = 0 ,
\end{equation}
where $\bnabla^\rT \bi{T}$ is the force per unit volume, now represented by a row matrix. Therefore, provided $\bnabla\epsilon = 0$ and $\varrho= 0$ on the surface, the total force and torque acting on $\Omega$, (\ref{t:force}) and (\ref{t:torque}), can be calculated by using the Maxwell stress tensor (\ref{TE}). In fact, by virtue of (\ref{equilibrium}), the $\bi{T}_0 + \rho E^2 (\partial_\rho \epsilon) \mathbf{\mathds{1}}/2$ term in (\ref{stress}) is a uniform pressure over the surface that makes no contributions.

Because of the $\epsilon$ discontinuity, when the integration surface is the interface between two media, this simplification is of no help and the Cauchy stress tensor (\ref{stress}) must be used in (\ref{t:force}) and (\ref{t:torque}). However, provided that the body of interest is housed in an homogeneous medium, the calculation of (\ref{t:force}) and (\ref{t:torque}) can be still simplified by observing that, since $\bi{T}\bi{n}$ is continuous through the surface, it does not matter if the Cauchy stress tensor in the body (\ref{stress}) or the Maxwell stress tensor in the medium (\ref{TE}) are used. In addition, the integration surface can be any; as long as it encloses the body, but not additional free charges.

\section{Finite element analysis}
To calculate the Maxwell stress tensor, we used a commercial finite element analysis software \cite{COMSOL} to solve numerically the Laplace equation for an inhomogeneous medium,
\begin{equation}\label{field:eq}
 \bnabla^\rT (\epsilon \bnabla \phi) = 0 ,
\end{equation}
where $\phi$ is the field potential in the coil-magnet gap -- the cylindrical domain shown in Fig.\ \ref{toy-2}. In (\ref{field:eq}), the dielectric constant is $8.854$~pF/m in the vacuum and $53.4$~pF/m in the ceramic rings. % glass-ceramic MACOR: eps_r = 6.03
Eventually, the electric field and displacement are
\numparts\begin{eqnarray}
 \bi{E} &= &-\bnabla \phi \\
 \bi{D} &= &\epsilon\bi{E} .
\end{eqnarray}\endnumparts

Dirichlet boundary conditions were specified on the domain boundaries. In particular, $\phi=0$~V on the surface of the magnetic circuit whereas, owing to the electrical current and the relevant ohmic potential drops, the electric potential of the coil surface is assigned as shown in Fig.~\ref{toy}. In the figure, the current gets into the coil from the top-outer winding, which is set to the $RI^2$ potential, and it gets out from the top-inner winding, which is set to the zero potential. The current reversal does not change the electrostatic interaction, but the grounding reversal -- that is, the setting of the potential of the top-inner winding to $RI^2$ and of the top-outer winding to zero potential -- was also considered. The mesh, of about $7.9 \times 10^6$ tetrahedral elements, was the result of successive refinements; the relative numerical tolerance was set to $10^{-12}$.

Once the electric field has been calculated, the Maxwell stress tensor was obtained by the application of (\ref{TE}). To calculate the force acting on the coil, we integrated $\bi{T}_E$ over the torus $\Sigma$ -- having rectangular cross-section and cutting midway the magnetic gap -- shown in Fig.~\ref{toy-2}; the $\bi{E}$ value on $\Sigma$ was the external one-sided limit of the electric field. The next section motivates these choices.

\begin{figure}
\centering
\includegraphics[width=40mm]{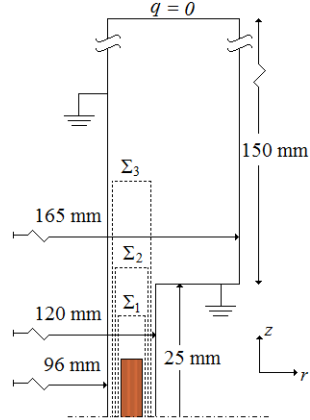}
\caption{Radial section of the upper half of the cylindrical capacitor used for the assessment of the finite element analysis. The gap accommodates three tori -- $\Sigma_1$, $\Sigma_2$, and $\Sigma_3$ -- enclosing the movable armature. The dimensions of the moving armature (brown-coloured) are: inner radius 105~mm, outer radius 111~mm, height 20~mm.}\label{demo}
\end{figure}

\begin{figure}[b]
\centering
\includegraphics[width=64mm]{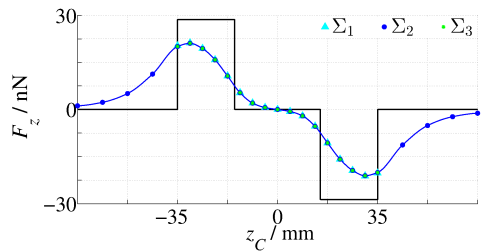}
\caption{Vertical force acting on the movable armature of the capacitor shown in Fig.~\ref{demo} {\it vs.} the positions $z_C$ of its centre; $z_C=0$~mm is a centred armature. The force was calculated by integrating $\bi{T}_E$ over the three tori shown in Fig.~\ref{demo}. The stepped line is the force calculated by the application of (\ref{e-force}).}\label{test}
\end{figure}

\subsection{Assessment of the finite element analysis}
The concepts outlined in section \ref{eltforces} and the capabilities of the finite element analysis were investigated by calculating the vertical force, $F_z$, acting on the movable armature of the cylindrical capacitor shown in Fig.~\ref{demo}. In order to test the independence of $F_z$ on the integration surface, the magnetic gap has been increased to 24~mm and three nested tori, enclosing the movable armature, have been accommodated in it. In figure~\ref{demo}, they are shown by dashed lines. Dirichlet boundary conditions were specified on the armature surfaces, where $\phi = 10$~V, and on the inner and outer surfaces of the field domain, where $\phi = 0$~V. On the top and bottom surfaces it was specified a zero charge, to ensure a field domain having open ends. The force was calculated by integrating $\bi{T}_E$ over the each torus. The results are compared in Fig.~\ref{test}; they do not show appreciable differences. The finite element calculation smooths the sharp steps originated by the application of (\ref{e-force}), as expected. It also shows that (\ref{e-force}) overestimates the maximum force. Tests were also made to verify the influence of the sharpness of the domain boundaries, which was found negligible.

The finite element software integrates $\bi{T}_E$ by using both the (internal and external to the integration surface) one-sided limits of the electric field. The relevant force values are given in table \ref{table:demo}: owing to the $\bi{E}$ continuity, they are expected to be the same. The best match is found for the $\Sigma_2$ surface. In addition, the excellent agreement between the force values relevant to $\Sigma_{2+}$ (external one-sided limit) and $\Sigma_{3-}$ (internal one-sided limit) suggests that the most accurate $\bi{E}$ values are in the region between $\Sigma_2$ and $\Sigma_3$. Hence, in the finite element model of the coil/magnet system, we located the integration surface midway between the coil and magnet, as shown in Fig.~\ref{toy-2}, and used  the external one-sided limit.

\begin{table}
\caption{\label{table:demo} Vertical force acting on the movable armature of the cylindrical capacitor shown in Fig.~\ref{demo}; $z_C$ is the vertical coordinate of the armature centre and $z_C=0$~mm is a centred armature. The force has been calculated by integrating $\bi{T}_E$ over the three tori shown in Fig.~\ref{demo}, $\Sigma_1$, $\Sigma_2$, and $\Sigma_3$; both the one-sided limits are given -- internal, $\Sigma_{i-}$, and external, $\Sigma_{i+}$, to the integration surface. The standard deviation, $u_{\rm diff}$, of the force differences -- which differences are expected to be null -- are also given.}
\begin{indented}
\item[]\begin{tabular}{ccccccc}
\br
 $z_C$/mm &\multicolumn{6}{c}{$F_z$/nN} \\
          & \multicolumn{1}{c}{$\Sigma_{1-}$} & \multicolumn{1}{c}{$\Sigma_{1+}$}
          & \multicolumn{1}{c}{$\Sigma_{2-}$} & \multicolumn{1}{c}{$\Sigma_{2+}$}
          & \multicolumn{1}{c}{$\Sigma_{3-}$} & \multicolumn{1}{c}{$\Sigma_{3+}$} \\
\hline \\
\multicolumn{1}{r}{$-35.000$} & \multicolumn{1}{r}{$20.313$} & \multicolumn{1}{r}{$20.276$} & \multicolumn{1}{r}{$20.066$} & \multicolumn{1}{r}{$20.057$} & \multicolumn{1}{r}{$20.059$} & \multicolumn{1}{r}{$20.082$}\\
\multicolumn{1}{r}{$-30.625$} & \multicolumn{1}{r}{$21.057$} & \multicolumn{1}{r}{$21.164$} & \multicolumn{1}{r}{$21.100$} & \multicolumn{1}{r}{$21.096$} & \multicolumn{1}{r}{$21.095$} & \multicolumn{1}{r}{$21.113$}\\
\multicolumn{1}{r}{$-26.250$} & \multicolumn{1}{r}{$19.324$} & \multicolumn{1}{r}{$19.428$} & \multicolumn{1}{r}{$19.407$} & \multicolumn{1}{r}{$19.405$} & \multicolumn{1}{r}{$19.406$} & \multicolumn{1}{r}{$19.420$}\\
\multicolumn{1}{r}{$-21.875$} & \multicolumn{1}{r}{$15.830$} & \multicolumn{1}{r}{$15.843$} & \multicolumn{1}{r}{$15.830$} & \multicolumn{1}{r}{$15.828$} & \multicolumn{1}{r}{$15.829$} & \multicolumn{1}{r}{$15.839$}\\
\multicolumn{1}{r}{$-17.500$} & \multicolumn{1}{r}{$10.806$} & \multicolumn{1}{r}{$10.667$} & \multicolumn{1}{r}{$10.666$} & \multicolumn{1}{r}{$10.666$} & \multicolumn{1}{r}{$10.666$} & \multicolumn{1}{r}{$10.671$}\\
\multicolumn{1}{r}{$-13.125$} & \multicolumn{1}{r}{$5.409$} & \multicolumn{1}{r}{$5.337$} & \multicolumn{1}{r}{$5.347$} & \multicolumn{1}{r}{$5.346$} & \multicolumn{1}{r}{$5.349$} & \multicolumn{1}{r}{$5.349$}\\
\multicolumn{1}{r}{$-8.750$} & \multicolumn{1}{r}{$2.024$} & \multicolumn{1}{r}{$2.044$} & \multicolumn{1}{r}{$2.039$} & \multicolumn{1}{r}{$2.038$} & \multicolumn{1}{r}{$2.040$} & \multicolumn{1}{r}{$2.038$}\\
\multicolumn{1}{r}{$-4.375$} & \multicolumn{1}{r}{$0.571$} & \multicolumn{1}{r}{$0.627$} & \multicolumn{1}{r}{$0.626$} & \multicolumn{1}{r}{$0.626$} & \multicolumn{1}{r}{$0.626$} & \multicolumn{1}{r}{$0.626$}\\
\multicolumn{1}{r}{$0.000$} & \multicolumn{1}{r}{$-0.026$} & \multicolumn{1}{r}{$-0.010$} & \multicolumn{1}{r}{$0.000$} & \multicolumn{1}{r}{$0.001$} & \multicolumn{1}{r}{$-0.000$} & \multicolumn{1}{r}{$0.000$}\\
\multicolumn{1}{r}{$4.375$} & \multicolumn{1}{r}{$-0.632$} & \multicolumn{1}{r}{$-0.635$} & \multicolumn{1}{r}{$-0.627$} & \multicolumn{1}{r}{$-0.627$} & \multicolumn{1}{r}{$-0.627$} & \multicolumn{1}{r}{$-0.627$}\\
\multicolumn{1}{r}{$8.750$} & \multicolumn{1}{r}{$-2.014$} & \multicolumn{1}{r}{$-2.049$} & \multicolumn{1}{r}{$-2.038$} & \multicolumn{1}{r}{$-2.038$} & \multicolumn{1}{r}{$-2.039$} & \multicolumn{1}{r}{$-2.038$}\\
\multicolumn{1}{r}{$13.125$} & \multicolumn{1}{r}{$-5.371$} & \multicolumn{1}{r}{$-5.346$} & \multicolumn{1}{r}{$-5.346$} & \multicolumn{1}{r}{$-5.347$} & \multicolumn{1}{r}{$-5.348$} & \multicolumn{1}{r}{$-5.347$}\\
\multicolumn{1}{r}{$17.500$} & \multicolumn{1}{r}{$-10.814$} & \multicolumn{1}{r}{$-10.671$} & \multicolumn{1}{r}{$-10.665$} & \multicolumn{1}{r}{$-10.665$} & \multicolumn{1}{r}{$-10.665$} & \multicolumn{1}{r}{$-10.671$}\\
\multicolumn{1}{r}{$21.875$} & \multicolumn{1}{r}{$-15.910$} & \multicolumn{1}{r}{$-15.835$} & \multicolumn{1}{r}{$-15.830$} & \multicolumn{1}{r}{$-15.829$} & \multicolumn{1}{r}{$-15.829$} & \multicolumn{1}{r}{$-15.840$}\\
\multicolumn{1}{r}{$26.250$} & \multicolumn{1}{r}{$-19.263$} & \multicolumn{1}{r}{$-19.425$} & \multicolumn{1}{r}{$-19.406$} & \multicolumn{1}{r}{$-19.403$} & \multicolumn{1}{r}{$-19.403$} & \multicolumn{1}{r}{$-19.420$}\\
\multicolumn{1}{r}{$30.625$} & \multicolumn{1}{r}{$-20.942$} & \multicolumn{1}{r}{$-21.183$} & \multicolumn{1}{r}{$-21.100$} & \multicolumn{1}{r}{$-21.095$} & \multicolumn{1}{r}{$-21.094$} & \multicolumn{1}{r}{$-21.114$}\\
\multicolumn{1}{r}{$35.000$} & \multicolumn{1}{r}{$-20.256$} & \multicolumn{1}{r}{$-20.267$} & \multicolumn{1}{r}{$-20.066$} & \multicolumn{1}{r}{$-20.056$} & \multicolumn{1}{r}{$-20.058$} & \multicolumn{1}{r}{$-20.082$}\\
%\multicolumn{1}{c}{$u_{\rm diff}$/nN} & \multicolumn{2}{c}{$0.101$} & \multicolumn{2}{c}{$0.004$} & \multicolumn{2}{c}{$0.013$}\\
\multicolumn{1}{c}{$u_{\rm diff}$/nN} & \multicolumn{6}{c}{$0.101~~~~~~~~0.078~~~~~~~~0.004~~~~~~~~0.001~~~~~~~~0.013$}\\
\br
\end{tabular}
\end{indented}
\end{table}

\begin{figure}
\centering
\includegraphics[width=50mm]{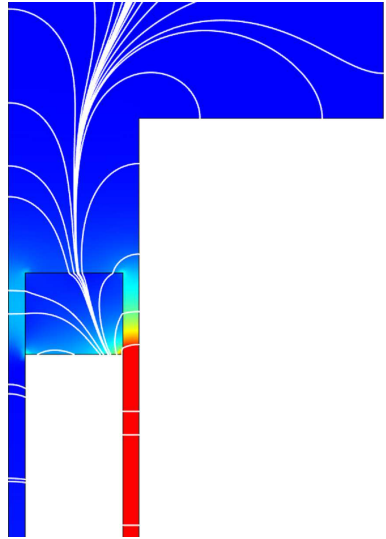}
\caption{Zoom of the radial section of the electric field in the magnetic gap; only part of the upper domain is shown. The colours indicate the field intensity, from $0$~V/m (blue) to $2.5$~kV/m (red). The field streamlines are also shown. The coil is in the centred, $z_C=0$~mm, position. The boundary conditions are given in Fig.~\ref{toy}.}\label{field:map}
\end{figure}

\newpage
\subsection{Results}
\begin{figure}
\centering
\includegraphics[width=64mm]{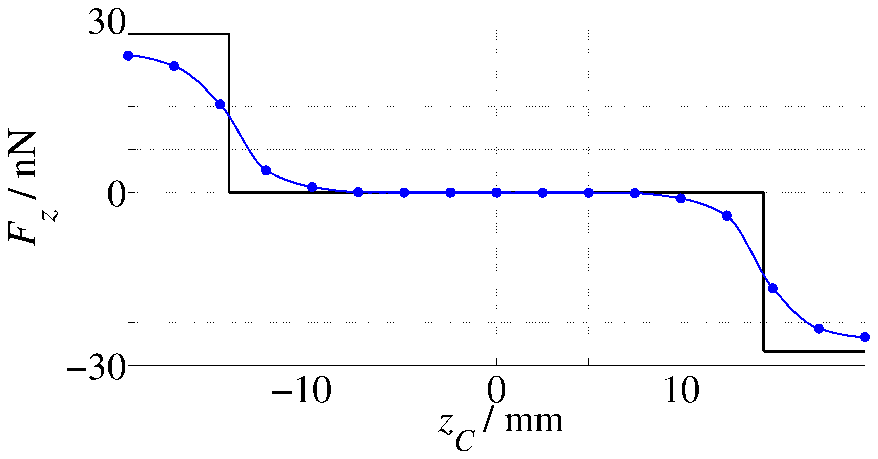}
\includegraphics[width=64mm]{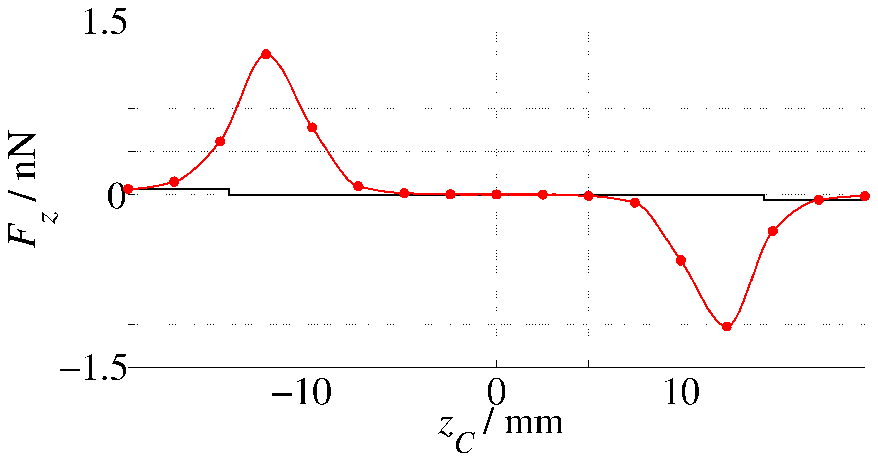}
\caption{Vertical force acting on the coaxial coil {\it vs.} the vertical position $z_C$ of its centre; when $z_C=0$~mm the coil is centred. The stepped lines are the forces calculated by the application of (\ref{e-force}). Left: the inner top-winding is set to the zero potential. Right: the outer top-winding is set to the zero potential.}\label{offset:vertical}
\includegraphics[width=64mm]{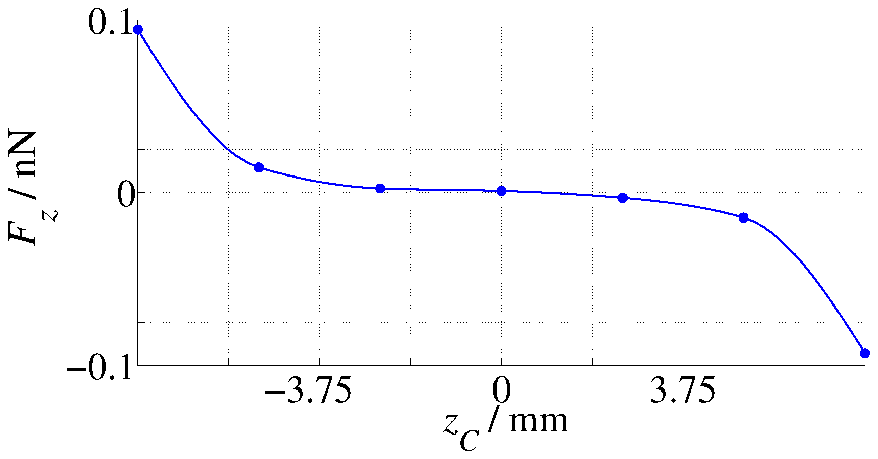}
\includegraphics[width=64mm]{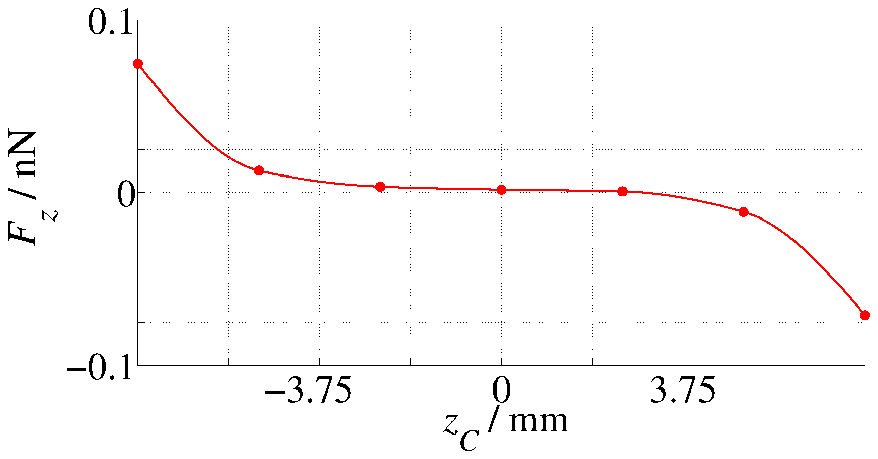}
\caption{Zoom of the vertical force for small coil displacements from the centred, $z_C=0$~mm, position. Left: the inner top-winding is set to the zero potential. Right: the outer top-winding is set to the zero potential.}\label{zoom}
\includegraphics[width=64mm]{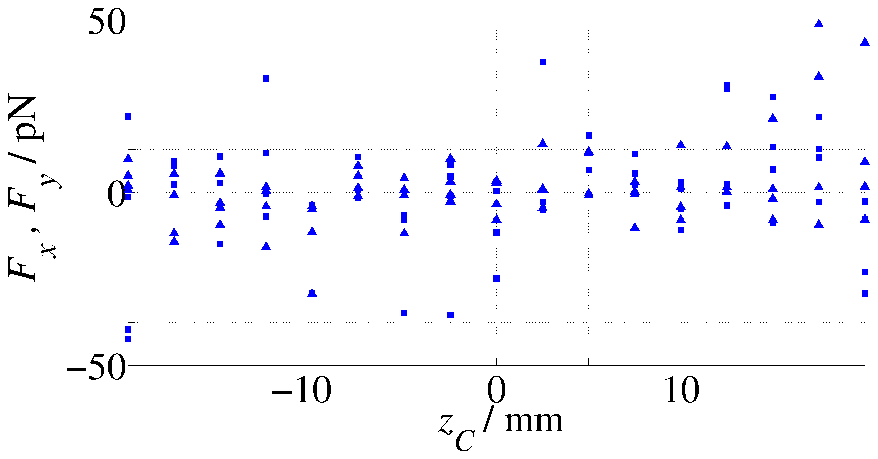}
\includegraphics[width=64mm]{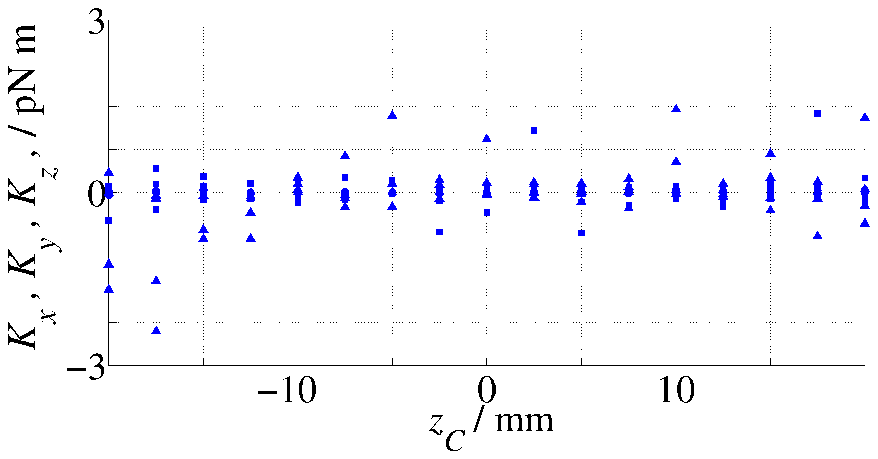}
\caption{Left: horizontal components of the force acting on the coaxial coil {\it vs.} the vertical position $z_C$ of its centre. Right: components of the torque about the centre acting on the coaxial coil {\it vs.} the vertical position $z_C$ of its centre. When $z_C=0$~mm the coil is centred. All the values calculated -- with both the inner and outer top-windings set to the zero potential and with both the inner- and outer-field calculations of $\bi{T}_E$ -- are shown.}\label{horizontal}
\includegraphics[width=64mm]{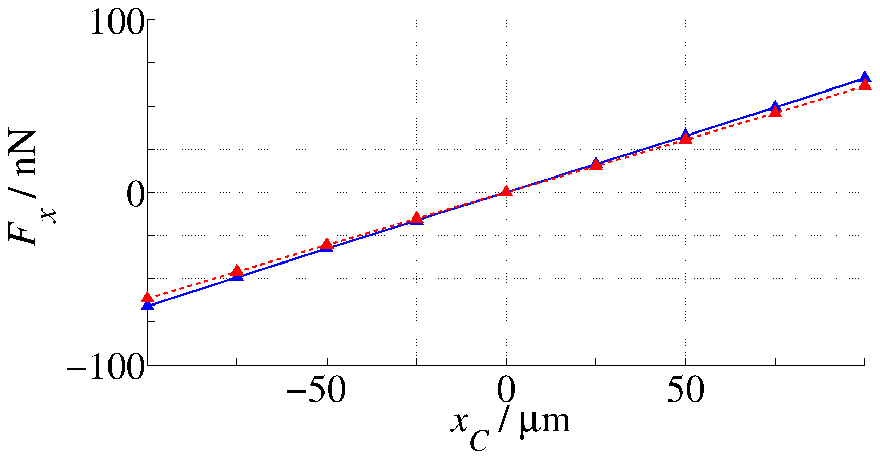}
\includegraphics[width=64mm]{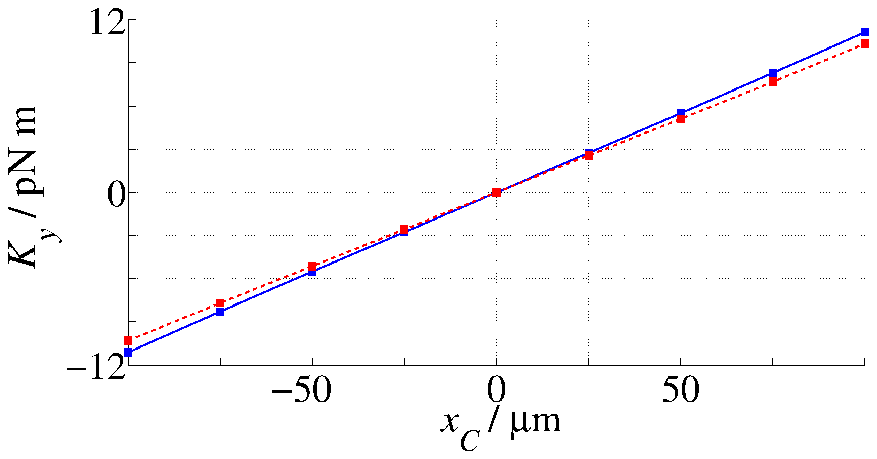}
\caption{Horizontal components of the force (left) and torque about the centre (right) {\it vs.} the radial offset $x_C$ of the coil. The coil centre is in the horizontal symmetry-plane of the magnet; the offset is along the $x$ axis. The $y$ force component and the $x$ torque components are null. Solid (blue) line: the inner top-windings has been set to the zero potential; dashed (red) line: the outer top-windings has been set to the zero potential.}\label{offset:radial}
\end{figure}

The two dimensional map of the the electric field in the magnetic gap -- for a coaxial and centred assembling of the coil-magnet system -- is shown in Fig.~\ref{field:map}. The vertical force was calculated by integrating the Maxwell stress tensor on the torus -- having a rectangular cross-section and embedding the whole coil -- shown in Fig.~\ref{toy-2}. The force has been calculated with different vertical positions of the coil centre and with both the inner and outer top windings set to the zero potential; the results are shown in Fig.~\ref{offset:vertical} together with the force obtained by the application of (\ref{e-force}).

Figure~\ref{zoom} shows a zoom of $F_z$ for small coil displacements; the origin is a stable equilibrium point having an elastic constant equal to $-1.1$~pN/mm. Since the centring rings eliminates -- from the viewpoint of the electric field -- the pole shoe, the inner capacitance is independent of the vertical coil-position and, when they are the outer coil windings to get the null potential, $F_z$ is always nearly zero. The minuscule raises at the coil entrance and exit from the magnetic gap is due to the polarisation of the coil supporting-rings.

Owing to the system symmetry, the horizontal component of the electrostatic force is null; the values obtained by the finite element analysis are shown in Fig.\ \ref{horizontal}. The standard deviation of these data is $\pm~20$~pN. For the same symmetry reason, also the torque about the coil centre is null; the values obtained by the finite element analysis are shown in Fig.~\ref{horizontal}. The standard deviation is $\pm~0.4$~pN~m.

Eventually, we calculated the force acting on an off-centre coil; the results are shown in Figs.~\ref{offset:radial}. The coaxial assembly is an unstable equilibrium position having an elastic constant equal to 0.66~nN/$\mu$m. The torque about the coil centre is $0.11$~pN~m/$\mu$m; this value indicates that the centre of application of the force is 0.17~mm above the centre.

\section{Conclusions}
The watt balance operation requires that no external force acts on the coil, apart from that due to the interaction between the electric current and the magnetic field. However, since the coil resistance raises the coil potential with respect to that of the magnet, stray capacitances induce electrostatic forces. The METAS watt balance having been taken as a starting point of our simplified model, we reported about a finite element analysis aimed at quantifying these electrostatic interactions. {\colr Up to now, these forces have been assumed irrelevant. Our study shows that a finite element analysis has adequate accuracy for investigating their effect and that the no-effect assumption was indeed correct.}

{\colr Stray capacitances affect also the moving-mode operation: charge and discharge currents induced by the capacitance variations influence the measured voltage between the coils ends. Furthermore, the electrostatic forces acting on the moving coil can induce unwanted velocity components. These effects should be negligible, but they deserve future investigations nevertheless.

The magnetic equivalent of the electrostatic forces is related to the coil inductance and displays itself as a dependence of the magnetic field on the coil current and position. Since it is embedded in a detailed calculation of the magnetic forces acting on the coil, the calculation of the relevant parasitic forces requires a huge effort both from the theoretical and numerical viewpoints. It is a complex magnetostatic problem that includes the simulation of the permanent magnet, of the magnetic circuit, and of their response to the coil current. Once the field in the magnetic gap is on hand, one can proceed by evaluating the relevant the Maxwell stress tensor and by integrating it over a closed surface embedding the coil. This paper outlined, in a much simpler framework, the general strategy to cope with this problem.}

{\colr To help the metrologists performing watt-balance experiments, we summarise here the main results of our investigation.} By symmetry reasons and by neglecting the potential gradients on the coil surface, no electrostatic force acts on a coaxial coil placed in the magnet centre. From the electrostatic viewpoint, this is a stable equilibrium point with respect to vertical displacements and an unstable equilibrium point with respect to horizontal displacements. When misplacements are considered once at a time, the elastic constants are $-1.1$~pN/mm (vertical) and 0.66~nN/$\mu$m (radial). The potential gradients make the application point of this elastic force misplacement about 0.17~mm above the coil centre.

The vertical elastic constant is so small that, also in the case of a millimetre misplacement, the relevant bias can be neglected. The effect of the force and torque component in the horizontal plane, $F_h$ and $K_h$, can be examined as follows. As long as the magnetic force and mass weight are coaxial and pass through the coil centre of mass, there is no need for the coil to be in a given position. Therefore, if the balance is so aligned as to make it insensitive to the exchange of the mass weight for the coil force, we can assume that $F_h$ and $K_h$ are counteracted by opposite magnetic force and torque nullifying them and making the total force vertical and passing through the coil centre. In this case, the relevant measurement equation is \cite{Robinson:2012b}
\begin{equation}
 mgu_z = \cU I \left(1 + \frac{F_h}{mg}\frac{u_h}{u_z} + \frac{K_h}{mg}\frac{\omega_h}{u_z} \right) ,
\end{equation}
where $u_h$ and $\omega_h$ are the radial components of the velocity and angular-velocity, respectively, $mg=5$~N and, in the case of 1~mm offset between the field and coil axes, $F_h/(mg) \approx 1.3 \times 10^{-7}$ and $K_h/(mg) \approx 22 \times 10^{-12}$~m. Therefore, the relevant constraint on the coil motion are irrelevant.

In addition, it must be noted that the weighing is carried out by offsetting the balance by 0.5~kg and by adding the magnetic forces generated by equal and opposite currents in the coil with the 1~kg mass on and off the pan. Therefore, provided that the coil grounding is not changed in the current reversal, the electrostatic contribution to the total force cancels in the sum.

{\colr Design expedients to remove the electrostatic forces also exist. The filling of the magnetic gap with a non-magnetic metal to create a toroidal cavity of rectangular cross section increases the system symmetry thus reducing the vertical force-component. A second solution is to add an electrostatic shield around the coil, e.g., by winding a additional single-layer coils which are earthed at one point.}

\ack
This work was jointly funded by the European Metrology Research Programme (EMRP) participating countries within the European Association of National Metrology Institutes (EURAMET) and the European Union. We thanks A Eichenberger and H Baumann who let us take the METAS watt balance as a practical example for this investigation.

\end{document}